OPEN ACCESS

EDITED BY
Alfio Maurizio Bonanno,
Osservatorio Astrofisico di Catania
(INAF), Italy

REVIEWED BY
Enrico Maria Nicola Corsaro,
Osservatorio Astrofisico di Catania
(INAF), Italy
Alberto Vecchiato,
Osservatorio Astrofisico di Torino (INAF),
Italy

*CORRESPONDENCE
Tobias C. Hinse,
✉ toch@cp3.sdu.dk

RECEIVED 08 July 2023
ACCEPTED 17 August 2023
PUBLISHED 12 September 2023

CITATION
Hinse TC, Dorch BF, Occhionero LVT and
Holck JP (2023), How Tycho Brahe's
recordings in 1572 support SN 1572 as a
type I(a) supernova.
*Front. Astron. Space Sci.* 10:1255481.
doi: 10.3389/fspas.2023.1255481

# How Tycho Brahe's recordings in 1572 support SN 1572 as a type I(a) supernova

Tobias C. Hinse [1,2]*, Bertil F. Dorch [1,2], Lars V. T. Occhionero [3] and Jakob P. Holck [1]

[1]University Library of Southern Denmark, Odense, Denmark, [2]Department of Physics, Chemistry and Pharmacy, University of Southern Denmark, Odense, Denmark, [3]Department of Danish History of Astronomy, Kroppedal Museum, Taastrup, Denmark

The 450th anniversary of the discovery of the SN 1572 supernova event was celebrated in 2022. A closer look at the historical development of the field of supernova astronomy reveals the scientific importance of Tycho Brahe's 1572 observations of this "new star." In their quest to learn more about the new type of stellar explosion and subsequent evolution, the initial protagonists in this field (Baader and Zwicky among others) gradually turned their attention to the final remnant state of these supernova events. Since the remnant object thought to be associated with the extragalactic supernova event was found to be very dim, the focus quickly shifted toward nearby galactic events. It is at this point where Tycho Brahe's observations played an important and often overlooked role in the context of the development of stellar evolution as a scientific field. Tycho Brahe's meticulous and detailed recordings of the change in brightness of the new star not only allowed modern astronomers to classify SN 1572 as a supernova event but also helped them pinpoint the exact astrometric location of SN 1572. These findings helped to empirically link extragalactic supernova events to nearby past supernova remnants in the Milky Way. This enabled subsequent observations allowing further characterization. Transforming the historical recordings to a standardized photometric system also allowed the classification of SN 1572 as a type I supernova event.

KEYWORDS

history and philosophy of astronomy, stars: individual: SN 1572, B Cassiopeia, Tycho's star, supernovae: SN 1572, ISM: supernova remnants: SN 1572, stars: mass-loss, X-rays: general

## 1 Introduction

The year 2022 marks the 450th anniversary of the discovery of a guest, or new star, mysteriously appearing suddenly on the northern hemisphere within the constellation of Cassiopeia at a position that was known to be devoid of any known fixed star.

Records of the first sightings of a new star were performed by an abbot in Messina on the Island of Sicily (Italy) (Stephenson and Green, 2002) and W. Schuler in Wittenberg (Germany) (Baade, 1945) on the morning of 6 November 1572 (Julian calendar; see the following for details about the difference between the Julian and Gregorian calendar).

The discovery of the new star is credited to Tycho Brahe (born 14 December 1546, died 24 October 1601) who witnessed the new appearance of a bright star on the evening of 11

November 1572 (Julian date). The difference of 5 days between the first sightings/records by observers in Italy/Germany and Tycho's observation is likely explained due to bad weather in Denmark/Skåne (southern part of today's Sweden). No recordings of Tycho's health or wellbeing in the days prior to 11 November exist. Tycho's attribution or acknowledgment for the discovery of a new star is most likely on the merit of him publishing his recordings in the important 1573 publication "*De Nova Stella*" (Brahe, 1573). For scientists and philosophers of the early renaissance, this seminal publication is at the foundation of the later historical development of astronomy and the history of natural sciences. In "*De Nova Stella*," Tycho Brahe's discovery not only documented a change in the heavens beyond the orbit of the Moon but also marks the beginning of refuting Aristotle's idea (e.g., Aristotle's chief cosmological treatise "*De Caelo*") of the "unchanging heavens" and, therefore, was part of the early movement toward a shift in the then prevailing scientific paradigm.

The importance and subsequent dissemination of "*De Nova Stella*" eventually catapulted the Kingdom of Denmark, and Tycho Brahe himself, on the international arena of contemporary frontiers science and provided the initial financial seed for Tycho Brahe to pursue a life-long passion to carry out ground-breaking astronomical research and instrumentation.

Today, we know that the return of investment was surmounting and sparked the beginning of the important age of enlightenment (e.g., the "Great Age of Reasoning") in Western culture and European societies in general (Thoren, 2002).

For reasons of accuracy in dates given in this review, it may be interesting to note that the Gregorian calendar was introduced in 1582 and implemented in Denmark and Sweden on the 1st of March 1700, far later than 1572. At Tycho Brahe's time, the Julian calendar was used. Currently, the Julian calendar is 13 days behind the Gregorian calendar; however, in 1572, it would have been only 10 days behind. Therefore, Tycho's first recording of SN 1572, in the extrapolated Gregorian calendar, was on 21 November 1572.

Tycho Brahe's contribution was a systematic recording in time of the astrometric position[1] of the new star in an attempt to measure a daily (diurnal) parallax effect as a result of Earth's rotational motion. The result of these measurements would allow him to judge on the distance of the new star in relation to the distance to the Moon.

An interesting question to raise is what instrument did Brahe utilize when performing astrometric measurements of the new star? According to the work of Stephenson and Green (2002), the instrument depicted in Figure 1 was *used to measure the distance of the SN of AD 1572 from nearby stars*. No reference for this statement is offered. We have reasons to believe that this statement is not correct.

According to the work of Pedersen (1980) [pointing to the work of Thoren (1973)], two types of sighting instruments were utilized to measure the position of the new star and in which Brahe had one or the other involvement. The first instrument was a duplicate, but enlarged re-design of his first half-sextant (5 foot in length with an arc length of 30°) which he made use of in earlier times and as shown in Figure 1. According to the work of Thoren (1973), this instrument was *not* used by Tycho Brahe when observing the new star in November 1572. We quote from the work of Thoren (1973)

> When Tycho returned to Denmark in 1570, he left his half-sextant [Figure 1] behind in Augsburg as a gift. Accordingly, when he reached Skåne [aka. Scania, the southern province of today's Sweden] he set about duplicating the instrument, departing from the original pattern only to the extent of making the arc twice as large.

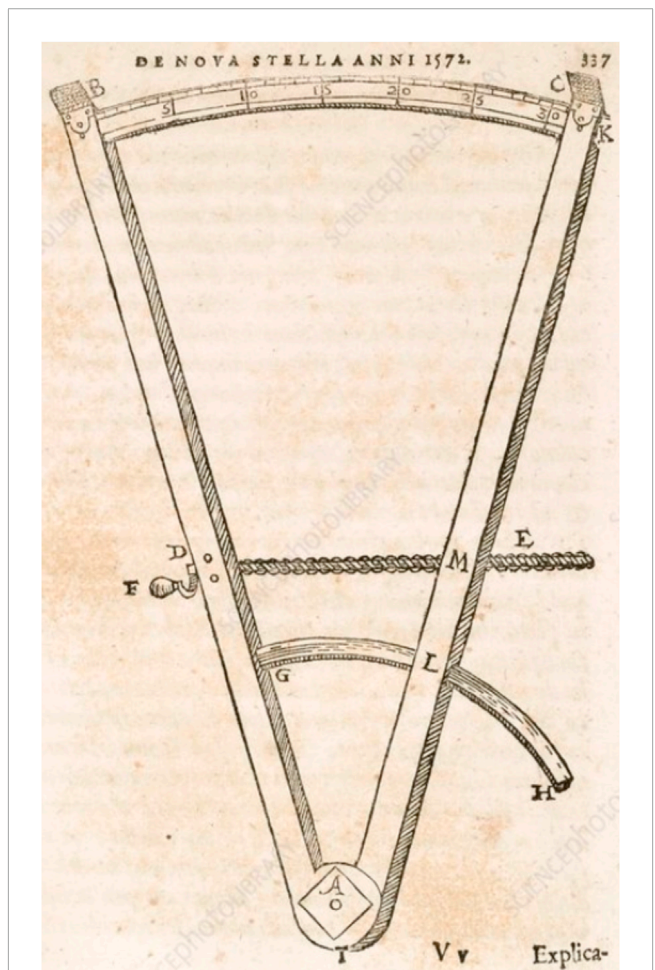

**FIGURE 1**
Predecessor instrument to the sextant used by Tycho Brahe. It is important to mention that this instrument was likely not used [as claimed otherwise by Stephenson and Green (2002)] to measure the position of the new star as observed in November 1572 and subsequent months. See text for details.

Thoren (1973) provided a reference to the work of Brahe and Dreyer (1913), Tomus V, p. 84–87) translated to English under the title *Tycho Brahe's discussion of his instruments and scientific work*.

---

[1] Two types of quantitative measurements were performed by Brahe: the position of the new star relative to three known (fixed) stars in the constellation of Cassiopeia and its position relative to the ecliptic coordinate system.

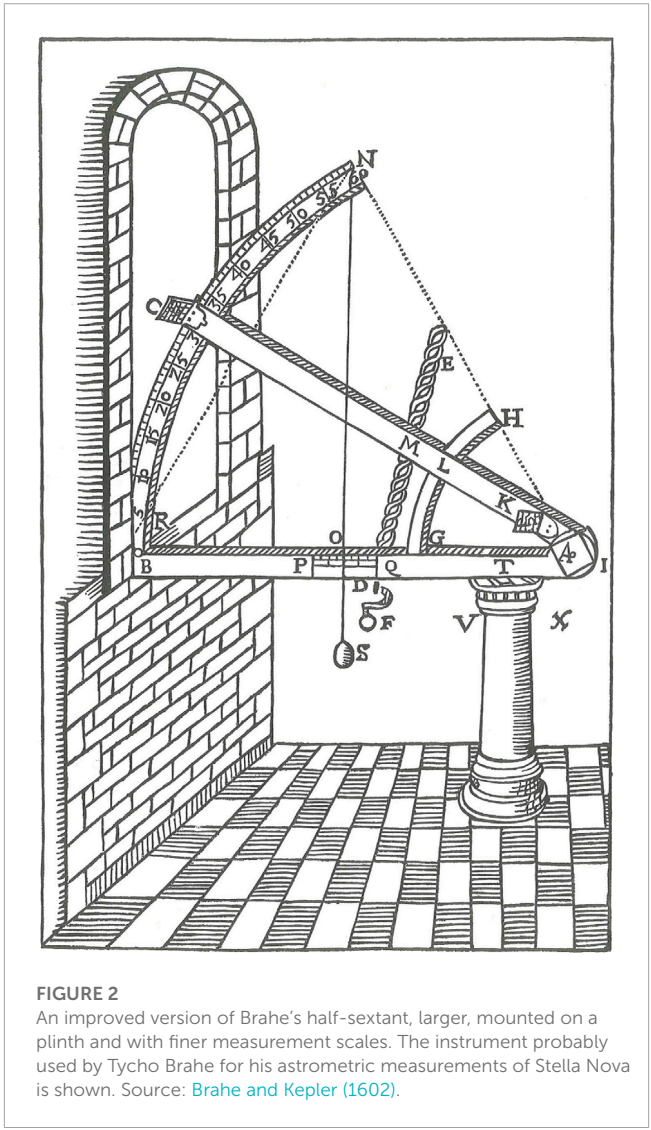

FIGURE 2
An improved version of Brahe's half-sextant, larger, mounted on a plinth and with finer measurement scales. The instrument probably used by Tycho Brahe for his astrometric measurements of Stella Nova is shown. Source: Brahe and Kepler (1602).

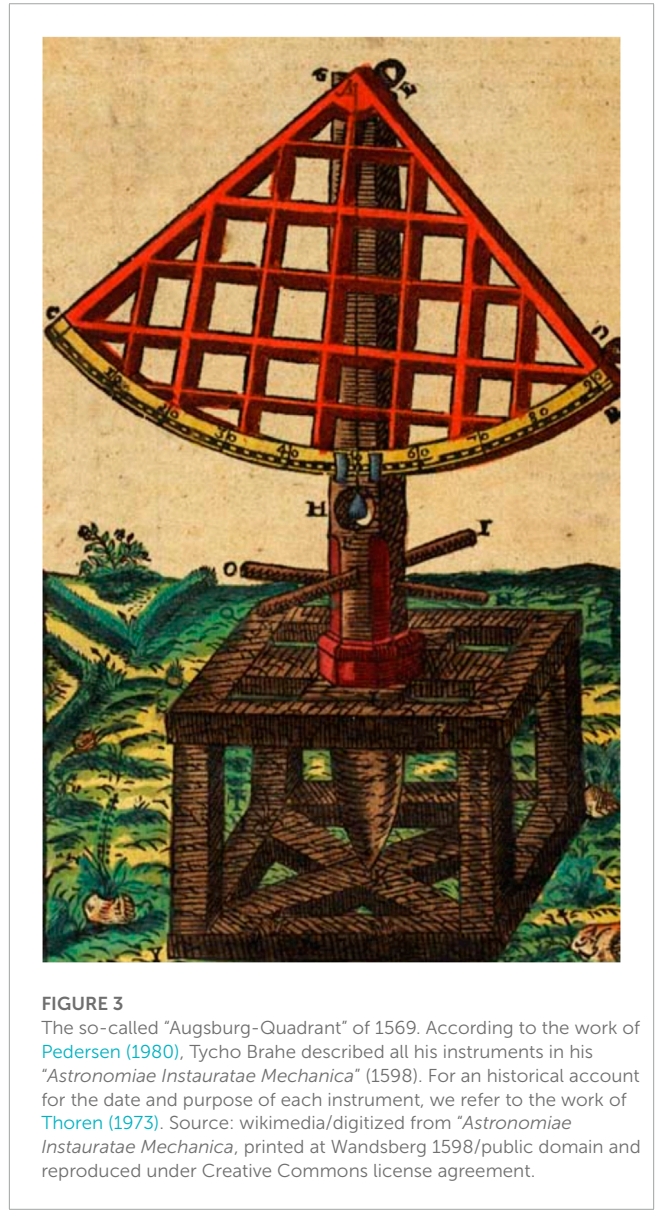

FIGURE 3
The so-called "Augsburg-Quadrant" of 1569. According to the work of Pedersen (1980), Tycho Brahe described all his instruments in his "*Astronomiae Instauratae Mechanica*" (1598). For an historical account for the date and purpose of each instrument, we refer to the work of Thoren (1973). Source: wikimedia/digitized from "*Astronomiae Instauratae Mechanica*, printed at Wandsberg 1598/public domain and reproduced under Creative Commons license agreement.

We, therefore, learn that Tycho Brahe embarked on improving his instrument (see Figure 1) in 1570 when returning to Denmark (Knutstorp Manor, Scania) after visiting contemporary fellow astronomers near the city of Augsburg, Germany. In the work of Brahe and Kepler (1602, p. 339), Tycho Brahe recollected how he sat waiting by the accurately tuned instrument, waiting to observe the new star. The instrument mentioned in the book is the one in Figure 2. This is, therefore, probably the instrument Tycho Brahe used to observe Stella Nova. The instrument is indeed an improved version of the older half-sextant, now mounted on a plinth by a window and with what seems to be more accurate measuring systems. This instrument is also depicted in the work of Brahe and Dreyer (1913), Tomus V, which Thoren (1973) used as reference. The motivation to improve his existing half-sextant might have its origin in Brahe's active involvement in the 1569 construction of the so-called "Augsburg-Quadrant" (see Figure 3) allowing him to improve his craftsman skills in designing a new instrument and test ideas for improvements at an early age in his life. According to the work of Pedersen (1980), the Augsburg-Quadrant was also used to measure the celestial position of the new star of 1572.

We, therefore, point out that Stephenson and Green (2002) likely are in error when referring to Figure 1 as the instrument used for astrometric measurements of the new star. The actual instrument used is probably the one depicted in Figure 2.

From Tycho Brahe's own measurement, the resulting non-detection of a diurnal parallax allowed him to conclude that the new star must be beyond the distance of the Moon. This measurement and a non-detection of any motion of the new star relative to the fixed stars (the five known classic planets did change position in the sky) allowed him to conclude that the new star must also be beyond the realm of the known planets and, therefore, belong to the realm of the fixed stars. This chain of argument and conclusion is worth paying attention to and marks a turning point in the history of science. In Tycho's own words in his *De Nova Stella* [Brahe (1573)] [from a translation taken from the work of Stephenson and Green (2002)],

*That it is neither in the orbit of Saturn, however, nor in that of Jupiter, nor in that of Mars, nor in that of any one of the other planets, is hence evident, since after the lapse of several months it has not advanced by its own motion a single minute from that place in which I first saw it … Hence this new star is located neither in the [elemental region[2]], below the Moon, nor in the orbits of the seven wandering stars, but in the eighth sphere, among the other fixed stars.*

For clarity of context, it is important to remember that at that time, the dominant belief system was a paradigm based on the Aristotelian/Ptolemaic system, where the realm of the stars (beyond the Moon and the well-known five classic planets) were never changing in position, static in relation to each other, constant in their appearance, and displayed no change in colors or brightness in the past, present, and in all eternity. At least this is what classic philosophers observed. Indisputable, the beginning of the end of this viewpoint, which was held dear in the minds of many for over two millennia, is marked with Tycho Brahe's detailed quantitative (astrometry) as well as qualitative (brightness and color) recordings of the new star, first noted by Tycho Brahe himself, on the evening of 11 November 1572 (Julian calendar). Figure 4 shows the drawing of the new star by Tycho Brahe in relation to several other stars in the constellation of Cassiopeia.

In light of the enormous impact of Brahe's contribution to the development of modern astronomy[3], it seems a daunting task to account for all the glory details of that development from a scientific history point of view; let alone all biographic details of T. Brahe's life. We will not dwell too much on various aspects of Tycho Brahe's scientific legacy. This would be beyond the scope of this review paper. We instead refer to both historic as well as modern and recent published monographs on the life and work of Tycho Brahe (Christianson, 2000; Thoren, 2002). However, with respect to one important point, we would like to emphasize the following which sets the framework of this review: Tycho Brahe—even in modern times—is best (if not only) known for his long-term, accurate, and precise (1 arc minute or less precision) measurements and recordings of the heavenly positions of stars and planets. His recordings of the new star in 1572, however—even at his own time—lend him less importance and fame in his subsequent contemporary track-record of scientific achievements. Apparently, not many fellow astronomers paid attention to his 1573 monograph. Only in the context of the development of scientific history and, especially, in the subsequent development of a more modern theory of the motion of the planets, the significance of Brahe's work gained momentum and importance and the proper recognition that Brahe is bestowed with today in the 'hall of fame' of important astronomers.

This review paper aims to cover two tasks. The first task is an attempt to provide a review of literature (see Appendix) that aimed at presenting historical data/recordings by Tycho Brahe and contemporary scientists that allowed a modern-day quantitative reconstruction of the time variation in apparent brightness—or in modern language, the light curve—of the "new star" as described by Tycho Brahe in "*De Stella Nova*" and later with some more details in "*Progymnasmata*" published posthumously in 1602 by Johannes Kepler. Interestingly and likely overlooked, the field of making use of historical recordings is currently still very much active and proves surprisingly fruitful. Recent focus on the use of data spanning two thousand years was applied to the temporal color variation of Betelgeuse and Antares (Neuhäuser et al., 2022, and references therein).

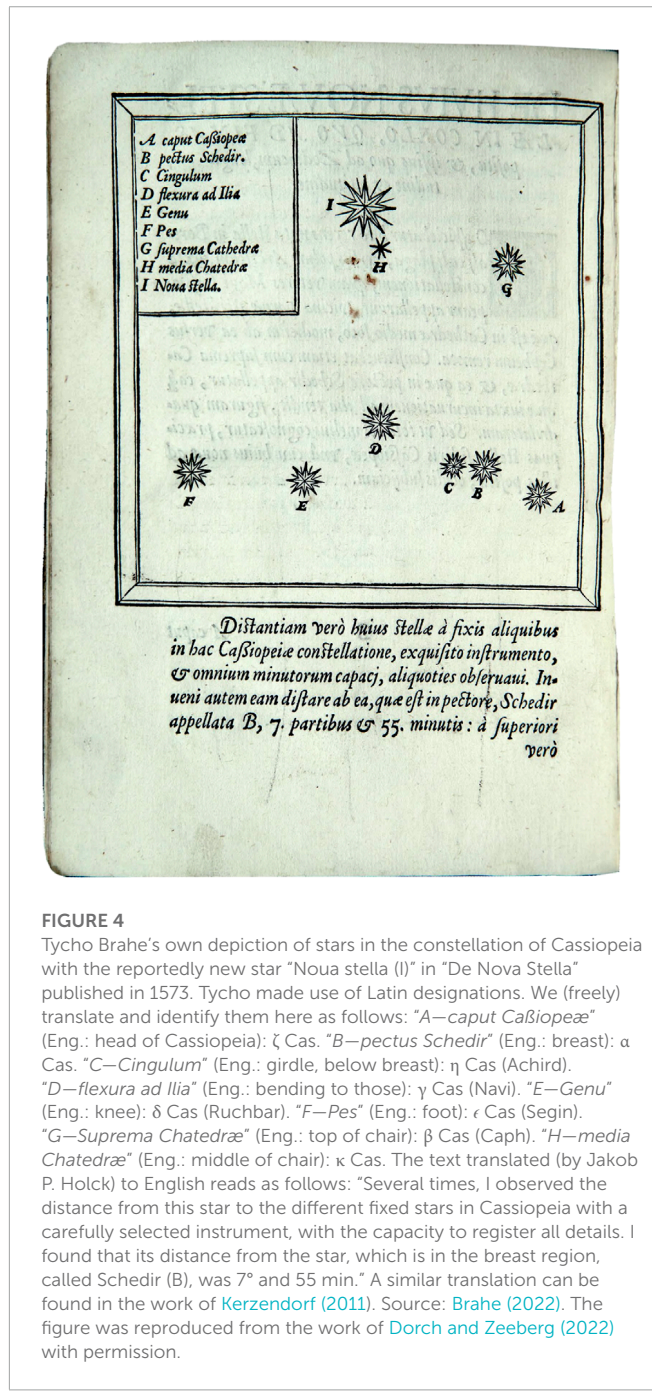

FIGURE 4
Tycho Brahe's own depiction of stars in the constellation of Cassiopeia with the reportedly new star "Noua stella (I)" in "De Nova Stella" published in 1573. Tycho made use of Latin designations. We (freely) translate and identify them here as follows: "*A—caput Caßiopeæ*" (Eng.: head of Cassiopeia): ζ Cas. "*B—pectus Schedir*" (Eng.: breast): α Cas. "*C—Cingulum*" (Eng.: girdle, below breast): η Cas (Achird). "*D—flexura ad Ilia*" (Eng.: bending to those): γ Cas (Navi). "*E—Genu*" (Eng.: knee): δ Cas (Ruchbar). "*F—Pes*" (Eng.: foot): ι Cas (Segin). "*G—Suprema Chatedræ*" (Eng.: top of chair): β Cas (Caph). "*H—media Chatedræ*" (Eng.: middle of chair): κ Cas. The text translated (by Jakob P. Holck) to English reads as follows: "Several times, I observed the distance from this star to the different fixed stars in Cassiopeia with a carefully selected instrument, with the capacity to register all details. I found that its distance from the star, which is in the breast region, called Schedir (B), was 7° and 55 min." A similar translation can be found in the work of Kerzendorf (2011). Source: Brahe (2022). The figure was reproduced from the work of Dorch and Zeeberg (2022) with permission.

The second task is the result of asking the question: *what is the origin of the terms "type I" and "type II" when classifying supernova events?* This question arises naturally when one examines in detail the work presented by Baade (1945), where, for the first time, the

---

light curve of the supernova of 1572 (SN 1572 or Tycho's new star) was reconstructed very carefully and with great precision from detailed historical records compiled by Brahe. The title of Baade's (1945) paper is "B Cassiopeiae as a supernova of type I." One interesting aspect of this paper is the use of the classification term "type I" *based on only photometric data*. We quote from the work of Baade (1945)the following: "*The recent recognition of two types of supernovae makes it desirable to decide whether the star was a supernova of type I or type II. The light curve of the nova, derived in the present paper, clearly indicates a supernova of type I.*" No reference to any spectroscopic data was given.

This is peculiar in several ways: *i)* given the fact that in a much earlier publication, Minkowski (1941) proposed two classes of supernovae events—"type I" and "type II"—based on spectroscopic data only and *ii)* Baade (1945) did not present any spectral evidence along with the reconstructed light curve with an opportunity to provide a reference to the work of Minkowski (1941). However, the latter part seems at that time a daunting/insurmountable task given the limited aperture of telescopes available at that time since several efforts to identify the remnant star of SN 1572 failed as a (conjectured) result of being very faint. Today, we know that the remnant of SN 1572 is not easily detectable in visible light. Obviously, and especially because spectroscopy is a "light-hungry" measurement, this prohibited the acquisition of any spectroscopic observations in the beginning of the 1900s. However, the curious reader will ultimately be stumbled upon this lack of accuracy in the classification of SN 1572.

As a result of a continued literature review, we encountered a similar inaccuracy already taking place historically in the work by Baade (1943), where the author reconstructed the light curve of the 1604 supernova (SN 1604 or Kepler's Supernova). Again, the supernova event was classified from the historic light curve as of type I and we quote from the introduction of the work of Baade (1943) the following: "*The light curve, derived in the present paper, shows that the star was a supernova of type I, which at maximum reached the apparent magnitude -2.2.*" Again, as was the case for SN 1572, also in this work, no spectroscopic data were presented, completely missing the opportunity to provide a spectroscopic classification according to the different chemical characteristics described in the work of Minkowski (1941).

This manuscript is structured as follows. In Section 2, we present a brief review of historical records of Tycho Brahe's supernova SN 1572 (B Cas) resulting in the construction of modern light curves. We attempt to provide some details of the original recordings in this. In this respect, the work by Stephenson and Green (2002) proves a great resource. Some discussion of recordings from Far East to Asia is given, but most recordings of scientific values were produced in Europe. In Section 3, we review the early use of the "type I" and "type II" classification and trace out the coining of the term "supernova." We also discuss the historical importance of three papers—mainly by F. Zwicky and W. Baade—that we deem necessary for the understanding of finding an answer to our question set out in the introduction. In Section 4, we present in chronological order of appearance three papers that adopt the newly introduced two-group classification (type I and type II). In Section 5, we take a more detailed look at Baade's (1945) paper with the aim to trace out his train of thoughts to classify Tycho's supernova SN 1572 as a type I supernova.

## 2 Part 1—historical records of SN 1572

The first to systematically compile historical records on the brightness change of SN 1572 was W. Baade. In his 1945 paper (Baade, 1945), he presented all recordings of sightings made by Tycho Brahe himself and contemporary observers in Europe and transformed the data to a modern magnitude system. In a footnote, a single reference was made to a sighting from Korea agreeing with what European observers saw in late 1572. Baade's motivation to dwell on historical records was a desire to classify Tycho's nova either as a type I or type II supernova. The motivation for this differentiation will be described later. The identification of SN 1572 as a supernova (versus a common nova) was already fairly established within the community.

The key to allow for this transformation is to be found in the knowledge of brightness of known celestial objects. Tycho and others used stellar brightness classification based on the classic magnitude system as depicted in Ptolemy's Almagest catalog, as well as bright planets such as Jupiter and Venus for their comparisons. However, historically, negative magnitudes for brighter objects like Jupiter and Venus were not yet introduced.

The seminal works by Tycho Brahe published in the work of Brahe (1573) containing a preliminary report and the more extensive presentation of the recording found in *Astronomiae instauratae progymnasmata* (*secunda pars*) (Brahe and Kepler, 1602) (written in Latin), which also includes records from other observers in Europe, serve as the main source for Baade's goal to derive a light curve. As noted by Baade (1945), it seems that only Tycho Brahe deemed it useful to record the brightness change from the beginning of the flare-up in early November 1572 until the star's disappearance around March 1574. In order to obtain some traction on how Tycho described his observations we reproduce some of the recordings as translated by Baade (1945).

> When first seen [11 November 1572] the nova outshone all fixed stars, Vega and Sirius included. It was even a little brighter than Jupiter, then rising at sunset, so that it equalled Venus when this planet shines in its maximum brightness.

> The nova stayed at nearly this same brightness through almost the whole of November. On clear days it was seen by many observers in full daylight, even at noontime, a distinction otherwise reserved to Venus only. At night it often shone through clouds which blotted out all other stars.

> In December it was about equal to Jupiter. In January [1573] it was a little fainter than Jupiter and surpassed considerably the brighter stars of the first class. In February and March it was as bright as the last-named group of stars. In April and May it was equal to the stars of the second magnitude.

> After a further decrease in June it reached the third magnitude in July and August, when it was closely equal to the brighter stars of Cassiopeia, which are assigned to *this magnitude.*

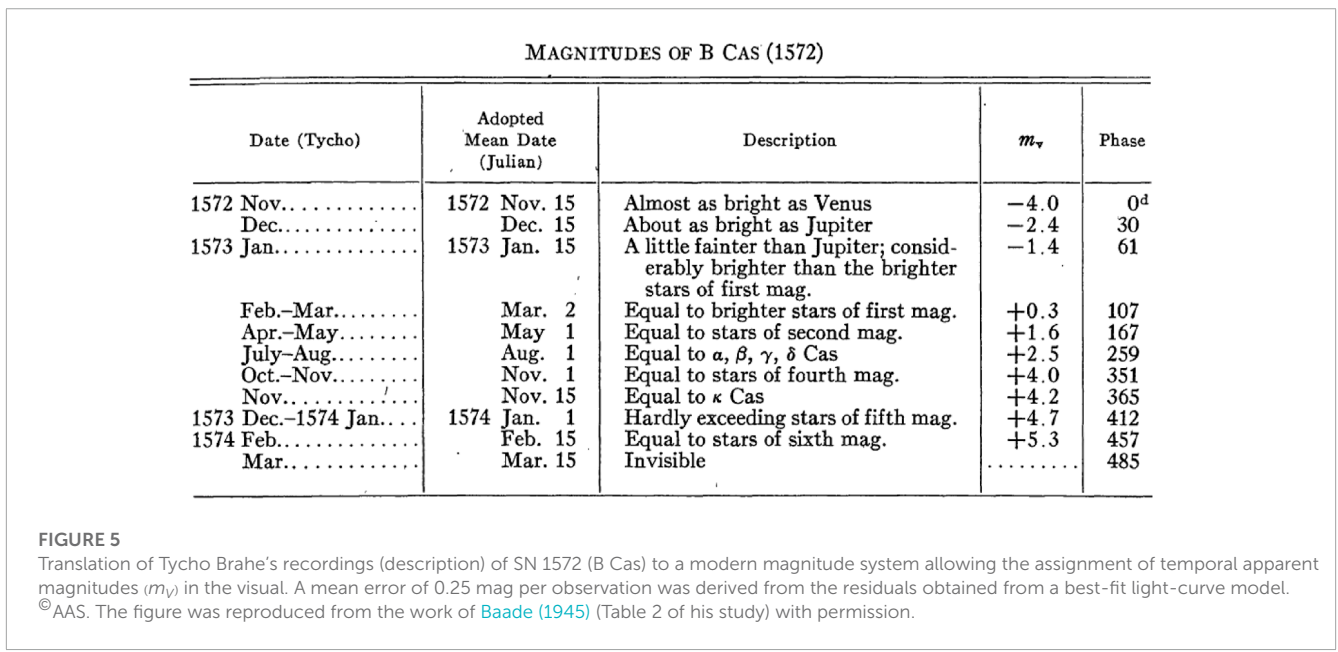

MAGNITUDES OF B CAS (1572)

| Date (Tycho) | Adopted Mean Date (Julian) | Description | $m_v$ | Phase |
|---|---|---|---|---|
| 1572 Nov. . . . . . . . . . . . | 1572 Nov. 15 | Almost as bright as Venus | −4.0 | 0$^d$ |
| Dec. . . . . . . . . . . . | Dec. 15 | About as bright as Jupiter | −2.4 | 30 |
| 1573 Jan. . . . . . . . . . . . | 1573 Jan. 15 | A little fainter than Jupiter; considerably brighter than the brighter stars of first mag. | −1.4 | 61 |
| Feb.–Mar. . . . . . . . . | Mar. 2 | Equal to brighter stars of first mag. | +0.3 | 107 |
| Apr.–May . . . . . . . . | May 1 | Equal to stars of second mag. | +1.6 | 167 |
| July–Aug. . . . . . . . . | Aug. 1 | Equal to $\alpha$, $\beta$, $\gamma$, $\delta$ Cas | +2.5 | 259 |
| Oct.–Nov. . . . . . . . . | Nov. 1 | Equal to stars of fourth mag. | +4.0 | 351 |
| Nov. . . . . . . . . !. . . | Nov. 15 | Equal to $\kappa$ Cas | +4.2 | 365 |
| 1573 Dec.–1574 Jan. . . . | 1574 Jan. 1 | Hardly exceeding stars of fifth mag. | +4.7 | 412 |
| 1574 Feb. . . . . . . . . . . . | Feb. 15 | Equal to stars of sixth mag. | +5.3 | 457 |
| Mar. . . . . . . . . . . . | Mar. 15 | Invisible | . . . . . . . . . | 485 |

FIGURE 5
Translation of Tycho Brahe's recordings (description) of SN 1572 (B Cas) to a modern magnitude system allowing the assignment of temporal apparent magnitudes ($m_v$) in the visual. A mean error of 0.25 mag per observation was derived from the residuals obtained from a best-fit light-curve model. ©AAS. The figure was reproduced from the work of Baade (1945) (Table 2 of his study) with permission.

*Continuing its decrease in September, it became equal to the stars of the fourth magnitude in October and November. During the month of November, in particular, it was so similar in brightness to the near-by eleventh star of Cassiopeia that it was difficult to decide which of the two was the brighter. At the end of 1573 and in January 1574, the nova hardly exceeded the stars of the fifth magnitude. In February it peached the stars of the sixth and faintest class. Finally in March it became so faint that it could not be seen any more.*

Baade's analysis is split into two parts. The first part focuses on the brightness change around the time of maximum and the second part on the slow brightness decline of the new star. During the maximum phase, brightness estimates of SN 1572 were naturally based on comparisons with either Jupiter or Venus. Known field stars were used as reference objects during the fading phase.

Baade carefully evaluates the sky position and brightness of both Jupiter and Venus at the time around late 1572 and early 1573. This is important because Tycho's recordings, at times, leave room for an ambiguous interpretation. For example, Tycho's phrasing "*equalled Venus when this planet shines in its maximum brightness.*" From this statement, it is not clear whether the observation refers to an actual (nightly) observation or an estimate based on past experience or from memory. In turns out that during November 1572, Venus was near or at maximum brightness approximately 130° from SN 1572 and could easily have served as a reference object.

In his brightness estimate and attention to details, Baade (1945) was careful toward the accuracy of Brahe's brightness estimate near maximum. Statements from both the preliminary report of 1573 (Brahe, 1573) and from the more detailed reporting in the work of Brahe and Kepler (1602) were compared with recordings made by contemporary observers from around Europe. Baade (1945) found that most observers agree that SN 1572 was distinctly fainter than Venus allowing one to conclude that Brahe's brightness estimate around maximum is somewhat very high. As a result, a final value near the maximum brightness of −4.0 was estimated. In Figure 5, we show a table of the translation of brightness reporting by Brahe to a modern magnitude system at various dates.

The appearance of SN 1572 around maximum brightness was also noted in other parts of the world. Following the work of Stephenson and Green (2002), the supernova SN 1572 was also observed by astronomers in the Far East. Sightings from China and Korea follow a then typical reporting style of reporting a single event with remarks on the size and appearance of the supernova without any referencing to objects of known brightness. Reports on the daytime observation of SN 1572 from China exist as well, confirming European sightings. The daytime flaring is plausible given the exceptional brightness and the circumpolar nature of the SN 1572.

From China, a total of five sightings exist. For example, the astronomical records from the *Mingshi* treatise indirectly mention a new star, and we quote from the work of Stephenson and Green (2002)

*There are also some (stars) which did not exist in ancient times but which exist now. Beside (pang) Cexing there is a guest star. During the first year of the Wanli reign period [= AD 1573] this newly appeared (chu). At first (xian) it was large; now (jin) it is small.*

Today, the star *Cexing* is identified with a star in the constellation of Cassiopeia allowing an inference of the relative position of the new star in relation to *Cexing*. Since the aforementioned quote was written in present tense, even at the time of fainting (small), the new star remained fixed (beside *Cexing*).

From Korea, two recordings exist. No recordings exist from Japanese observers. For further details on historical recordings from Southeast Asia, we refer interested readers to the work of Stephenson and Green (2002).

In the second part, Baade (1945) provided details in assigning modern apparent magnitudes to the various by-eye brightness estimates during the decline phase throughout 1573 and until the complete disappearance of SN 1572 in early 1574. For most parts, Tycho Brahe compares the brightness with stars of known magnitudes according to the Almagest catalog. In Figure 6, we show one of five tables by Baade (1945) with stars of class 3 in the Ptolemy magnitude system and the corresponding transformation to the modern Harvard system.

| Star | Ptolemy's Mag. | Harv. R.P. | Star | Ptolemy's Mag. | Harv. R.P. |
|------|------|------|------|------|------|
| α Cas............ | 3 | 2.47 | γ Cas............ | 3–2 | 2.25 |
| β Cas............ | 3 | 2.42 | δ Cas............ | 3 | 2.80 |

FIGURE 6
Magnitude of stars in the Cassiopeia constellation in the classic Ptolemy (stars of magnitude 3) and the revised standard Harvard photometric system. Tycho Brahe compared the brightness of SN 1572 with these stars during the time period of July–August 1573. ©AAS. The figure was reproduced from the work of Baade (1945) with permission.

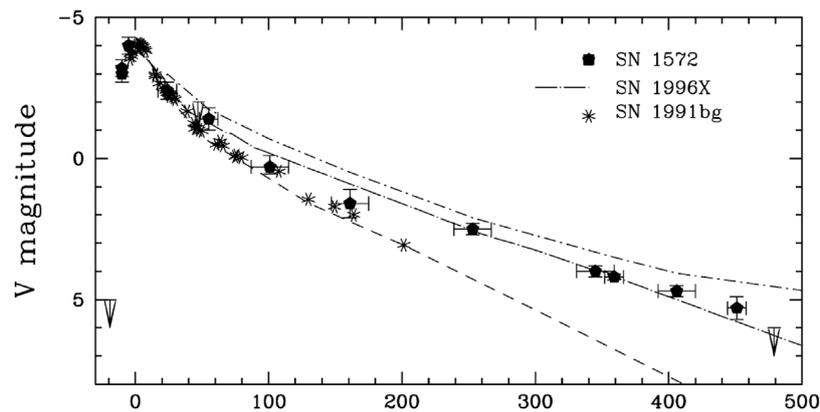

FIGURE 7
Modern light curve of SN 1572 as derived by Ruiz-Lapuente (2004) using largely Tycho's original observations. The total time window is around 510 days. The so-called stretch factor was found to be s 0.9 indicative of a SN Ia event. ©AAS. The figure was reproduced from the work of Ruiz-Lapuente (2004) with permission.

Baade emphasized that the reporting of Tycho Brahe strictly adheres to Ptolemy's magnitude system for field star comparisons. This approach by Brahe ensures inherent consistency or homogeneity for all reporting and, thus, accuracy in the transformation to modern magnitudes. However, some element of uncertainty, as pointed out by Stephenson and Green (2002), remains for the observation period around February–March 1573. Remarkably, and as a result of the strict adherence to the Ptolemy star class system, Tycho deemed it *not necessary* to include Sirius (α CMa; Ptolemy stars of magnitude 1) for his brightness comparison. From Figure 5, Tycho writes *equal to or brighter stars of first magnitude* and, thus, introduces some degree of imprecision. The exact meaning is unclear. The inclusion of Sirius would otherwise be a natural choice of reference in the transition from using Jupiter to the brightest field stars (Ptolemy stars of magnitude 1), especially given the angular distance between SN 1572 and Sirius, although low on the sky (atmospheric absorption would likely account for −0.5 mag), was less than the distance to Venus in late 1572 and early 1573. From the preceding discussion, it appears that the part of the light curve covering the early months of 1573 might be in error by around 0.5 mag.

However, following the work of Stephenson and Green (2002), one attempt to interpret Tycho Brahe's observation to also include Sirius as a reference object was pointed out by Kilburn (2001) who revisited , at that time, a newly discovered star atlas—*Uranographia Britannica* (published in 1750)—by the English astronomer John Bevis. Without specifying his sources, Bevis provided a summary of Tycho Brahe's observation and freely introduced an allusion to Sirius. From the work of Stephenson and Green (2002), we quote the wording by John Bevis.

However, from Tycho Brahe's own recordings in *De Stella Nova* (Brahe, 1573) or *Progymnasmata* (Brahe and Kepler, 1602), no reference to Sirius was ever made except the reference to stars of first magnitude.

Several other authors attempted to derive a light curve of SN 1572 from Tycho's historical recordings (Clark and Stephenson, 1977; Pskovskii, 1978; Schaefer, 1996; Stephenson and Green, 2002). The most recent derivation, and likely most accurate with attention to great detail, is given by Ruiz-Lapuente (2004) largely based on Tycho's recordings as presented in Baade (1945). The author carefully evaluated realistic uncertainties for each observation and applied extinction corrections. A useful and hitherto overlooked observation, dated on 11 November 1572, was quoted with a recording by Jerónimo Munñoz (?–1592; compiled from his monograph *Libro del nuevo cometa*) who was a professor of Hebrew and Mathematics at the University of Valencia and later of Astrology at the University of Salamanca. Interestingly, Munñoz also measured the diurnal parallax of SN 1572 and confirmed the null-result of Tycho Brahe. In Figure 7, we show the light curve of SN 1572 as recovered by Ruiz-Lapuente (2004).

| NGC | Type | $m-M$ | Obs. $m_{max}$ | Extrapol. $m_{max}$ | $M_{max}$ | Remarks |
|---|---|---|---|---|---|---|
| 224 | Sb | 22.2 | 7.2 | ........ | −15.0 | S Andr |
| 1003 | Sc | 26.8 | 12.8 | ........ | 14.0 | |
| 2535 | SBc | ...... | ...... | 14.7 | ........ | |
| 2608 | SBc | ...... | ...... | 11.0: | ........ | |
| 2841 | Sb | ...... | ...... | ...... | ........ | Only one observation |
| 4157 | Sc | ...... | ...... | 14.4 | ........ | |
| IC 4182 | Sc | 24.8 | 8.2 | ........ | 16.6 | |
| 4273 | Sc | 26.7 | 14.4 | ........ | 12.3 | Virgo cluster |
| 4303 | SBc | 26.7 | ...... | 12.8 | 13.9 | Virgo cluster |
| 4321, No. 1 | Sc | 26.7 | ...... | 11.9 | 14.8 | Virgo cluster |
| 4321, No. 2 | Sc | 26.7 | ...... | ...... | ........ | Virgo cluster; only one observation |
| 4424 | SBb pec. | 26.7 | ...... | 11.1 | 15.6 | Virgo cluster |
| 4486 | Eo | 26.7 | ...... | 12.0 | 14.7 | Virgo cluster |
| 4527 | Sc | 26.7 | ...... | 13.0 | 13.7 | Virgo cluster |
| 5236 | Sc | 24.8 | ...... | ...... | ........ | Maximum doubtful |
| 5253 | Irr. | ...... | 8.0 | ...... | ........ | Z Cent |
| 5457 | Sc | 23.8 | ...... | ...... | ........ | SS U Ma; maximum doubtful |
| 6946 | Sc | 25.3 | ...... | 12.9 | −12.4 | |
| Mean | ......... | ...... | ...... | ...... | −14.3 | |

FIGURE 8
Final photometric values of various supernova events as presented in the work of Baade (1938). For the "Rosetta Stone" supernova IC 4182, both the distance modulus ($m-M$) and the (extrapolated) peak brightness (Obs. $m_{max}$) were determined and, thus, served as a standard supernova. The type column refers to the type of host galaxy. ©AAS. The figure was reproduced from the work of Baade (1938) with permission.

## 3 Part 2—early use of type I and type II classification

### 3.1 Coining the term supernova

During the 1920s, and substantially in the 1930s with the steady increase of observational evidence, the idea of the existence of a special class of novae gradually entered the stage of reality in the minds of Humason, K. Lundmark, F. Zwicky, and W. Baade among others. Formally, according to ADS, the first-time reference to the term "super-novae" was given in Baade and Zwicky (1934b) entitled "*On Super-novae*" and published in January 1934. It appears that preference is toward the work of Baade and Zwicky (1934c) and Baade and Zwicky (1934a) (both[4] published in May 1934) for the first-time use of the new terminology. In a short conference contribution, Osterbrock (2001) offered a historic trace back on the origin of the term "super-novae." According to Osterbrock, the term was mentioned first time orally by F. Zwicky at an American Physical Society meeting at Stanford in December 1933. The conference paper was later published in 1934 (Baade and Zwicky, 1934d) and according to Osterbrock is a condensation of the two 1934 publications. Furthermore, Osterbrock highlighted the use of alternative designation for the new class of novae and quoted several bibliographic sources without proper referencing. We attempted to find the proper references without any luck. Here, we repeat a few alternatives. In 1920, K. Lundmark (1920) described the new class as "*giant novae,*" and later, Lundmark (1923) described them as "*much more luminous novae.*" Further designation was "*exceptional novae*" (Hubble 1929), and W. Baade referred to them as "*Hauptnovae*" (chief novae). In a review paper, Zwicky (1940) highlighted in a footnote the comment

> Baade and I first introduced the term "supernovae" in seminars and in a lecture course on astrophysics at the California Institute of Technology in 1931.

However, following the work of Osterbrock (2001), the word "*super-Novae*" was first published by K. Lundmark in a December 1932 publication. The originality is questioned. Lundmark visited California in 1932–1933, and it is, therefore, argued for that he likely picked up the term "super-novae" from attending various seminars given his extended duration of stay. In conclusion, according to the work of Osterbrock (2001), it was W. Baade and F. Zwicky who coined "super-novae." In later times, the writing style without a hyphen was predominantly used as is done in modern times (Koenig, 2005).

---

4   The two publications are important in another respect, and we refer to the work of Burrows (2015) for a good review.

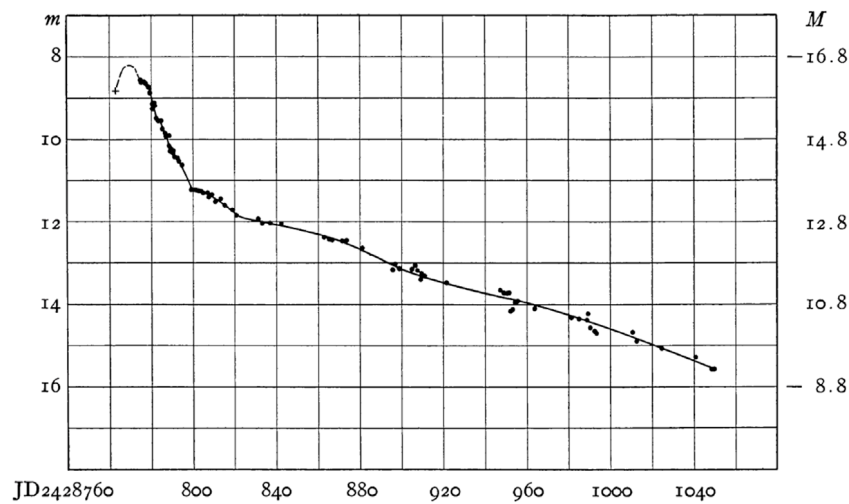

**FIGURE 9**
Light curve of the supernova IC 4182 as observed between August 1937 and June 1938 as presented in Baade and Zwicky (1938). The left scale designates photographic magnitudes and matches most closely observations through the modern Bessel/Johnson *B* filter pass band (Pierce and Jacoby, 1995). The right scale designates the absolute magnitude as a result of the known distance modulus to IC 4182. The single pre-maximum measurement was obtained by chance (see text for details). ©AAS. The figure was reproduced from the work of Baade and Zwicky (1938) with permission.

## 3.2 The Baade 1938 paper

In light of the steady development in understanding the nature of novae and novae-like events (Lundmark, 1923) the focus was shifted on the final state or remnants of novae and supernovae (Baade, 1938). In particular, the focus was on the end-stage result of past supernova events. Since the brightest novae were observed in extragalactic systems (see Figure 8), there was little hope to gain a deeper understanding due to their faintness. At that time and as a consequence of the limited ability of telescopes to gather light, attention was paid to past novae-like events in the local (Milky Way) galaxy. This point was raised by Baade (1938) dedicating the last section to the novae B Cassiopeia (SN 1572) and the Crab Nebula (SN 1054). Baade (1938) wrote

> *"Nothing is known at present about the final state of supernovae. Indeed, it would require a supernova in our own galaxy to obtain this information. Fortunately, we know two objects in our galaxy which very probably have been supernovae and which may provide an answer to our question: B Cassiopeiae and the Crab nebula."*

However, knowledge on whether Tycho's (SN 1572) and Kepler's (1604) were to be classified as supernovae or not was still very much under debate during the mid-1930s. As toward the nature of the Crab Nebula (SN 1054), Baade (1938) provided a footnote pointing out that K. Lundmark examined historical observations from Japan "*as bright as Jupiter*" strongly indicating SN 1054 to belong to the class of supernova. We refer interested readers to the work of Lundmark (1939) for further details on historical recordings from observers in Japan. The last paragraph in the work of Baade (1938) gives an account of all the arguments in support of SN 1054 being a former supernova. The exact nature (common novae vs. supernova) of SN 1572 and SN 1604 was still unclear.

In a quest to derive a life-luminosity relation of nova and nova-like events, Zwicky (1936) wrote

> *"Assuming the validity of the life-luminosity relation (1), some interesting applications can be made. For instance, the view has been advanced that Tycho's Nova 1572 and perhaps Kepler's Nova 1604 were super-novae."*

From this, we learn that in a relative sense, more certainty was lent toward SN 1572 for being a supernova event. However, the amount of empirical evidence given was sparse at that time. According to the work of Baade (1938), the only indication for SN 1572 to be a supernova event is based on its "*unusual brightness at maximum*" (Baade, 1938), and in the following, we will outline the reasoning provided by Baade in his 1938 paper.

The authors noted that SN 1572 is often cited as a *possible* supernova event because of its unusual brightness at maximum. This observation poses an important constraint in the following discussion. The author described a chain of evidence-based reasoning arguing in favor for that SN 1572 was not a common nova. A key reference is made to an important study by Humason and Lundmark (1923) where spectral properties of all stars in a region around SN 1572 down to *V* = 14 mag were investigated in a quest to identify the remnant star in association with the SN 1572 nova event.

The result of this study was negative reporting nothing unusual in their measurements. As a side note [footnote in the work of Baade (1938)], one star as measured by Humason and Lundmark (1923), located in the very vicinity of Tycho's originally measured position of SN 1572 and, thus, considered as a good candidate for the remnant star of the SN 1572 event, exhibited an M-type spectrum. At that time, as a result from a different study, at least one other star (T Coronae Borealis), with an albeit questionable M-type spectrum, was associated with a nova event allowing a *possible* bridging between spectral properties and nova events. However, according to the work of Baade (1938), the bridging was excluded by Lundmark because of the unusual spectrum.

The importance of the null-result was first realized several years later when in 1938, Humason was able to demonstrate that 16 (without exception) former nova events had spectral properties similar to very early B- or O-type spectra. Since in 1922, no stars in the vicinity of SN 1572 brighter than $V = 14$ mag had early B- or O-type spectra, Baade (1938) concluded that if SN 1572 was an ordinary nova event, then it must be fainter than 14th magnitude. Given the observed peak brightness as reported by Tycho Brahe himself (or translation thereof to a modern magnitude scale) was around −4 to −5 mag, the amplitude of SN 1572 light variation must have been at least 18–19 mag.

This unusual large amplitude provides a strong argument for that SN 1572 was not a common (or ordinary) nova event. At that time in 1938, the *mean amplitude* of common or ordinary nova events was found to be around 9th mag. The large amplitude was further increased to more than 22 mag by observations carried out in 1937 by W. Baade himself (Baade, 1938) as a result of a magnitude-limited photographic search survey in the red and blue with a limiting magnitude of $V = 18$. It is important to note that this conclusion (SN 1572 being a supernova) is only possible with the help of Tycho's meticulous recording of the (peak) brightness change of the new star at the end of 1572. It is also important to note that this paper (Baade, 1938) is the first to pinpoint, based on observational evidence and proper reasoning, that SN 1572 likely was a supernova event. Finally, we should also note that Baade (1938) did not offer to classify SN 1572 as a type I or type II supernova. The only clear differentiation is made between common or ordinary novae and supernovae. An explicit classification of SN 1572 being a type I supernova event was first given in the work of Baade (1945) (B Cassiopeiae as a Supernova of Type I) where Baade reconstructed the light curve SN 1572 from historical records.

## 3.3 The "Rosetta Stone" supernova

In the work of Baade and Zwicky (1938), the authors described the discovery and light curves of two particular supernovae events of which one later will play an important part in the supernovae classification of both the Kepler (SN 1604) and Tycho (SN 1572) supernovae events. Therefore, this paper deserves some attention in the present discussion. The underlying photometric data were acquired as part of a dedicated sky-monitoring program searching for bright novae using a Schmidt telescope at the Palomar Observatory in California. The respective host galaxies were included in the survey as a result of their known distance readily established by other means at that time in the form of a measured distance modulus.

In the period 1937–1938, the authors recorded the change in brightness of a supernova (see Figure 9) in the irregular spiral galaxy (then called nebula) IC (Index Catalog) 4182 under favorable observing conditions. The galaxy itself was discovered by W. Baade 2 years earlier in 1936 and is relatively faint and free of interstellar extinction as a result of being located in a relatively void region of the sky outside of the galactic plane. The supernova itself is designated as IC 4182. Subsequently, the supernova IC 4182 was designated as SN 1937C (Pierce and Jacoby, 1995). However, for the remainder of this review, we continue using the original designation. A total of six telescopes of various apertures were used for the subsequent photometric follow-up observations (see Table 2 in the work of Baade and Zwicky (1938)).

It appears that some initial hesitation existed toward whether or not to include IC 4182 in the final observing list. This is because IC 4182 as a galaxy is relatively sparse in the number of host stars lowering the chance to observe an associated nova/supernova event. As often in science, chance encounters appear, and within a year or so the bright supernova IC 4182 was discovered. The apparent brightness at the time of discovery was unusually high enabling the opportunity to obtain both photometric and spectroscopic observations. Historically, the latter dataset was obtained by R. Minkowski forming an important and significant part of his seminal supernova classification paper (Minkowski, 1941) published in 1941. In a forthcoming section, we will learn that IC 4182 falls in a particular class of supernovae events, thus, forming a "*Rosetta Stone*" linking similar supernova events with typological classification based on photometric data only.

Unfortunately, the discovery of the outburst of IC 4182 was made at past brightness maximum rendering the practical inference of the true brightness maximum to be near-impossible. As discussed earlier, the maximum apparent brightness of a nova event is a first empirical clue toward identifying the event as a supernova. Again, also in this respect, W. Baade and F. Zwicky were blessed with a large bag of luck. A single and most crucial observation was provided by a fellow astronomer and others, F. Leutenegger, who happened to observe comet Finsler in 1937. By chance, IC 4182 was in the field of view of one of Leutenegger's photographic plates obtained at a time predating the earliest observations obtained by W. Baade and F. Zwicky. This allowed them a determination of the brightness on the ascending branch of the light curve from which an estimate of the maximum could be inferred. For interesting reading on additional photometric data from historic archives of IC 4182, we refer to the work of Pierce and Jacoby (1995).

## 3.4 The Minkowski 1941 paper

As mentioned earlier, the first use of classifying supernovae as of either type I or type II was given in a seminal publication by Minkowski (1941) published under the title "*Spectra of Supernovae.*" Historically, we are here likely learning the reporting by Rudolph Minkowski of results from the first dedicated and systematic spectroscopic sky-survey of supernovae events observed in extragalactic systems.

We quote from the work of Minkowski (1941)

> "Spectroscopic observations indicate at least two types of supernovae. Nine objects (represented by the supernovae in IC 4182 and in NGC 4636) form an extremely homogeneous group provisionally called 'type I'. The remaining five objects (represented by the supernova in NGC 4725) are distinctly different; they are provisionally designated as 'type II'."

Important to emphasize is the following point: the classification is solely based on properties from spectroscopic observations. In the following, we shall assume that the reader is familiar with some details of spectroscopy and related terminology. Although the statistical sample is relatively meager (nine of "type I" and five of "type II"), Minkowski already noted a greater variation for the group of "type II" supernova events. In his own words, he wrote pointing already at an early stage toward a rich set of various sub-classes and/or peculiar supernova events (Milisavljevic and Margutti, 2018). Minkowski even contemplated the introduction of a third group based on one single supernova event, and we quote

> "The individual differences in this group [type II] are large…"

> "…at least one object, the supernova in NGC 4559 [subsequently named SN 1941A and classified as a type II-L supernova using the modern classification scheme], may represent a third type or, possibly, an unusually bright ordinary nova."

The introduction of additional types of supernova events is later taken up again by Fritz Zwicky (refer here his 1963 or 65 paper) although his classification did not survive to modern times. Minkowski then moved on to describe the temporal variation of spectra for the two types of supernovae. He first described properties of type I supernovae spectra. With the exception of minor differences,

> "the spectrograms of all objects of type I are closely comparable at corresponding times after maxima."

Furthermore, the spectra exhibited a property that was even present

> "very wide emission bands"

> "…at the earliest premaximum stage hitherto observed… No significant transformation of the spectrum occurs near maximum."

In a later paragraph, Minkowski (1941) reports on further properties of type I supernova spectra. He wrote

> "No satisfactory explanation for the spectra of type I has been proposed. Two (O I) [single-ionized oxygen] bands of moderate width in the later spectra of the supernova in IC 4182 are the only features satisfactorily identified in any spectrum of type I. They are, at the same time, the only indication of the development of a nebular spectrum for any supernova."

From the work of Minkowski (1941), we recall that the nine supernovae events of "type I" were observed from 7 days before maximum and, thus, we obtain the picture of a homogeneous class of spectra for supernovae of type I. We point out the spectrum of a supernova observed in the extragalactic (nebula) system IC 4182. Later, this type I supernova event will play a major role for the classification of Tycho's supernova of 1572 (as well as Kepler's supernova of 1603).

Minkowski (1941) then proceeded and reported his findings describing supernovae of type II. Spectra for this class have been recorded from maximum and until 115 days thereafter. Minkowski wrote

> "Up to about a week after maximum, the spectrum is continuous and extends far into the ultraviolet, indicating a very high color temperature. Faint emission is suspected near $H\alpha$ [Strongest emission line in the Balmer series at 6563 A]. Thereafter, the continuous spectrum fades and becomes redder. Simultaneously, absorptions and broad emission bands are developed. The spectrum as a whole resembles that of normal novae in the transition stage, although the hydrogen bands are relatively faint and forbidden lines are either extremely faint or missing."

The reported faint detection of hydrogen is the modern defining hallmark of a type II supernova. For type I supernovae, all spectra were absent of hydrogen. Minkowski then proceeded to remark that while the spectra of a type II supernova event is fairly understood from the construction of synthetic spectra and the resemblance to common or normal novae, the spectra of type I events are still lacking a satisfactory explanation (we recall the quote "No satisfactory explanation for the spectra of type I has been proposed"). It is interesting to pay special attention to this remark as this is likely the first historic identification for the profound difference of a type I and type II supernova explicitly pointing out the profound difference manifested in the underlying detonation/explosion physics.

Finally, Minkowski (1941) established a relative temperature hierarchy between type I, type II supernovae, and ordinary/common novae. Minkowski wrote

> "As compared with normal novae, supernovae of type II show a considerably earlier type of spectrum at maximum, hence a higher surface temperature (order of 40,000°)."

Further at the very end, he reported

> "This suggests that the supernovae of type I have still higher surface temperature and higher level of excitation than either ordinary novae or supernovae of type II."

In other words, the temperature of type I supernova is found to be higher than for type II events which again is higher than for common or ordinary (normal) novae. This last remark by Minkowski makes it clear that supernovae of type I are the more energetic events compared to type II supernova events. Since, according to Minkowski (1941), type II events appear spectroscopically closer to the group of common or ordinary novae, and since novae are intrinsically less luminous compared to supernovae (this was already established before 1941), one can conclude that type I supernovae events must be more luminous events.

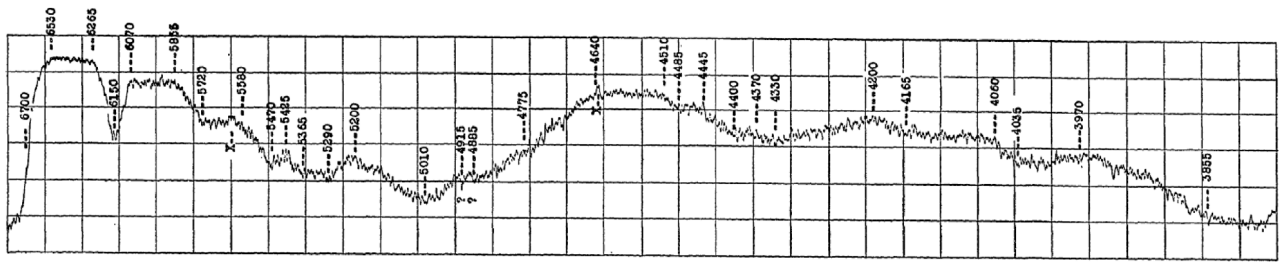

**FIGURE 10**
Spectrum of the "Rosetta Stone" supernova IC 4182 as observed on 31 August 1937. Nine days after maximum. Numerous other spectra were recorded as the supernova fainted. ©AAS. The figure was reproduced from the work of Minkowski (1939) (Figure 8) with permission.

A full presentation of data and analysis of all spectra for IC 4182 as observed in 1937–1938 was presented in the work of Minkowski (1939). Figure 10 shows a spectrum of IC 4182 observed some 9 days after maximum. In the quest to understand some spectral features several doppler-shifted lines were interpreted as being the result of an implosion of the progenitor object. This interpretation was in line with the neutron star hypothesis as suggested in Zwicky (1939), where Zwicky proposes the collapse of an ordinary star into a neutron star (Baade and Zwicky, 1934a; Baade and Zwicky, 1934d).

According to the work of Minkowski (1939), "*The absence of hydrogen and helium lines indicates a very high degree ionization; thus all the bands may be of unknown origin.*" Thus, here, we read the first-time report of the lack of hydrogen as a fundamental characteristic of a type I supernova.

## 3.5 Modern classification

The history of classifying supernova events starts in 1941 when Minkowski (1941) proposed two classes based on spectral observations. The first class provisionally coined "type I" showed no presence of hydrogen while the second class "type II" contained hydrogen (H$\alpha$ line detected). Details of subsequent spectral observations during the 1980s (both the number of supernova detection and data quality increased) made it necessary to introduce sub-classes according to additional chemical fingerprints. A modern classification system was reviewed by Turatto (2003), and Figure 11 shows a schematic of supernovae classification. Observations revealed that for most type I events an additional silicon (Si II at 6150Å) line was present resulting in the introduction of a type Ia supernova class. Historically, the original type I class introduced by Minkowski (1941) was later renamed type Ia. For those events with no evidence for the presence of silicon were further sub-classified according to the presence of helium. For supernovae with helium the class of type Ib and for helium-less objects the class of type Ic were introduced.

The astrophysical sites or cause of supernova events have also been identified. Two main mechanisms have been identified. For type Ia, the progenitor is the explosion of a white dwarf that accreted additional material from a nearby binary companion star eventually reaching some critical mass (Chandrasekhar mass limit) initiating a run-away carbon burning process driving a thermonuclear explosion. For the type Ib, Ic, and type II events, the progenitor is a massive star and the explosion is a result of a core collapse triggered at the very moment when the supportive radiation pressure ceases allowing a gravity-driven collapse into a neutron star or a black hole. The onset of both mechanisms mark the beginning of the end of the progenitor star's lifetime.

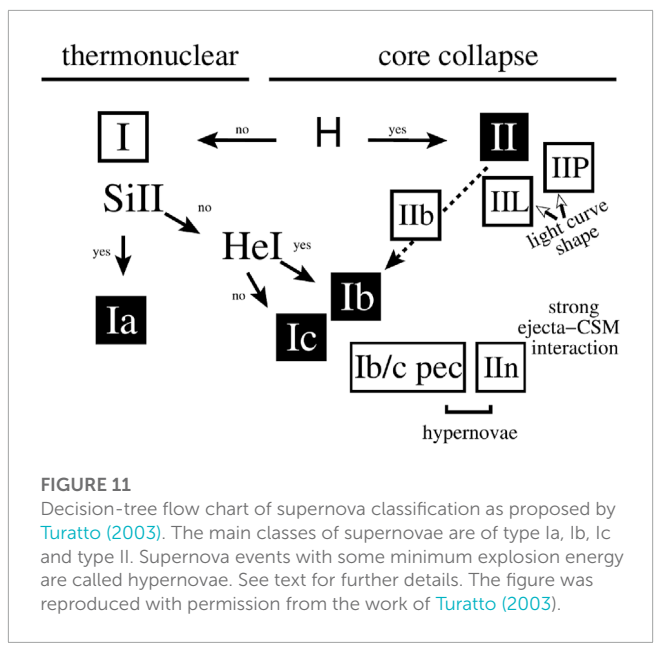

**FIGURE 11**
Decision-tree flow chart of supernova classification as proposed by Turatto (2003). The main classes of supernovae are of type Ia, Ib, Ic and type II. Supernova events with some minimum explosion energy are called hypernovae. See text for further details. The figure was reproduced with permission from the work of Turatto (2003).

The type Ia and II are further sub-classified according to kinematical properties or peculiarities or spectral-kinematic features. For type Ia SNe, we have the branch-normal sub-class, and for type II, we have the classes type IIP (plateau), IIL (linear), IIb, and IIn (narrow lines). A detailed discussion of these sub-classes is beyond the scope of this review and, thus, omitted, and we refer interested readers to Turatto (2003) for further details.

## 4 Three 1942 papers

In the year 1942, three papers adopted the newly introduced two-group classification of supernovae by Minkowski (1941). They deserve some in-depth attention with regards to the early use of the terms "type I" and "type II" since it appears that they are the only sources in the literature making use of the new classification terminology prior to W. Baade's classification of SN 1604 and SN 1572 as "type I" supernovae (Baade, 1943; 1945). It is remarkable to see how quickly the new finding of Minkowski (1941) is adopted by the main protagonist in novae/supernovae science and how old belief systems change as a reaction of the presentation of new evidence based on state-of-the-art observations.

## 4.1 Zwicky 1942 (April/July)

Zwicky (1942) published his second paper[5] concerning the occurrence frequency of supernovae entitled *On the frequency of supernovae. II*. He made use of the "type I" and "type II" terminology in the final paragraph where he wrote

> "*Another peculiar circumstance is the fact that those six supernovae among the twelve found during our initial patrol of the sky which were investigated spectroscopically were supernovae of what now is called "type I." This led us to a preliminary and incorrect conclusion that all supernovae might be of the same type. Some of the supernovae found later, such as the objects in NGC 5907 and 4725, proved to be of what Minkowski proposed to call "type II." These supernovae appear to be giant analogues of the common novae, and their spectra can be interpreted accordingly, while the interpretation of the spectra of supernovae of the type I has not yet been given. Supernovae of the type II, according to Baade, are, on the average, intrinsically fainter than supernovae of the type I; and they are therefore more difficult to discover, although they are probably more frequent than supernovae of the type I.*"

Around that time, Zwicky (1938) estimated one supernova per extragalactic nebula per 600 years. From this final remark, we learn from the work of Zwicky (1942) that prior to the identification (discovery) of two types of supernovae events, the main protagonists of supernovae research wrongly thought that only one type of supernova existed, though we have to remember that Zwicky published the paper as a single author only allowing us to indirectly infer that this view was shared by many in this field. Furthermore, the reference to the work of Baade when discussing "type II" supernovae is likely a reference to the work of Baade (1942) which is the second paper discussed in this section. Finally, as mentioned in Minkowski (1941), Zwicky reiterates that supernovae of "type II" have spectra that can be readily interpreted as they share similarities with common or ordinary novae events. However, a proper interpretation of spectra of "type I" is still missing. Historically, because of the lack of properly interpreting "type I" supernovae spectra, an increased interest is sparked to study galactic (within the Milky Way) supernova events. However, since these are rare events in the local Milky Way galaxy, the focus is on linking supernova in known extragalactic nebulosities with historic supernovae events. Tycho Brahe's observations of SN 1572 will play an important role in this linking.

---

5  April is the month stated at the end of the paper, and July is the ADS record.

## 4.2 Minkowski 1942 (May/September)

The paper by Minkowski (1942) carries the same title as the next publication by W. Baade. It seems almost as if the two authors (both were colleagues with offices in walking distance, with W. Baade at the California Institute of Technology and R. Minkowski at the Carnegie/Mount Wilson Observatory in Pasadena, California) fairly split the analysis of the Crab Nebula according to their respective scientific/observational expertise. Here, Minkowski (1942) presented details of spectroscopic observations of select regions in or near the Crab Nebula, and Baade (discussed in the next section) presented photometric observations. We chose to list Minkowski's (1942) paper first because of the May date-stamp presumably valid at the time of submission to the Astrophysical Journal, thus formally predating Baade's (1942) publication. However, the two papers have been published almost simultaneously in the Astrophysical Journal and are, thus, from a publication-time point of view indistinguishable.

We here again pay attention to the use of the terms "type I" and "type II." The first use by Minkowski is interesting, and we quote (omitting any reference to footnotes as present in the original print as they contain no relevant information)

> "*Little doubt remains that the Crab nebula is the remnant of the Chinese nova of A.D. 1054. This object was certainly a supernova; the records of its brightness indicate that it was a supernova of type I.*"

Surprisingly, Minkowski adopted Baade's differentiation between the two types of supernovae based on a measurement of brightness. We here obtain the impression that although, formally, two types of supernova have been empirically established based on spectral properties, Minkowski himself supported the association of "type I" supernovae with the more brighter family of supernovae. Further on in the text, Minkowski made use of "type I" classification as follows:

> "*For only one supernova of type I are there reliable data on the brightness before the outbreak. No star brighter than photographic magnitude 20.5 was present on earlier exposures of the nebula IC 4182 in the position of the supernova of 1937.*"

The classification of a supernova in the extragalactic system IC 4182 was already given by Minkowski in his seminal 1941 paper (Minkowski, 1941) being assigned "type I." The mentioning of IC 4182 here is used in order to provide observational evidence supporting the plausibility for a derived (from the mass–luminosity relation) absolute bolometric magnitude of −5 for the supernovae of 1054 at maximum. In the final paragraph, Minkowski then wrote

> "*If supernovae of type I are stars of mass greater than the critical mass $M_3$, then it is highly suggestive to assume that supernovae of type II are stars of mass smaller than $M_3$. Such an assumption does not meet any contradictory observational evidence.*"

*In its favor could be cited the fact that the frequency of supernovae of type II appears to be six times as great as that of supernovae of type I.In the absence of excess mass, a supernova of type II would not necessarily have to eject a considerable fraction of its mass. The nebula surrounding a supernova of type II should thus be fainter than that around a supernova of type I. This expectation is in general agreement with the fact that any nebula surrounding Tycho's nova of 1572, which was probably a supernova of type II, is certainly much fainter than either of two nebulae connected with supernovae of type I, namely, the Crab nebula and the nebula of Kepler's nova of 1604 recently found by Baade."*

In this text passage, besides learning about interesting thoughts with regards to the mass of stars before the breakout (detonation), we obtain Minkowski's opinion whether SN 1572 was of "type I" or "type II" based on the missing nebula "signature" of this supernova event. We have to remember that in 1942, no measurement of the maximum brightness of SN 1572 was yet known. The year 1942—with the pioneering study of Duyvendak (1942) and Mayall and Oort (1942)—marks the beginning of compiling and analyzing historic systematic records of past heavenly "guest stars" suddenly appearing on the sky. It appears very likely that the work by Duyvendak, Mayall and Oort inspired W. Baade to pursue similar work on SN 1604 (Baade, 1943) and SN 1572 (Baade, 1945) as we shall discuss soon. This surge in interest about information on the time evolution of the apparent brightness of historic supernovae events is explained in the quest to understand supernovae of "type I" in the nearby (Milky Way) galactic system. Up until 1942, "type I" supernovae were only observed in extragalactic systems.

## 4.3 Baade 1942 (June/September)

In the the work of Baade (1942) (*The Crab Nebula*), he presented photographs of the famous Crab Nebula and mainly carried out astrometric-kinematic measurements and calculations of nearby field stars and of the nebula itself. This contribution is a response to a recent identification (initiated by J. J. L. Duyvendak) of the nova of 1054 as the parent star resulting in the Crab Nebula. From historic data presented in the work of Mayall and Oort (1942) and Duyvendak (1942), Baade added the comment that their data implicitly allows the classification of the supernova of 1054 to be a "type I" event. However, of more interest is to read a footnote, and we quote

"*In the following discussion the term "supernova" always refers to a supernova of type I. Supernovae of type II, with luminosities intermediate between those of ordinary novae and supernovae of type I, appear to be closely related to the ordinary novae. In any case during an outburst that their present essentially the same phenomena as common novae.*"

Here, we are offered an alternative definition (strictly applicable only in the paper) of "type I" and "type II" supernovae based on the total power (luminosity) output.

## 5 Historic classification of Tycho's and Kepler's supernovae

In order to understand W. Baade's classification of Tycho's supernova of 1572 (SN 1572), it is useful to first outline his work on Kepler's supernova of 1604 (SN 1604) forming a necessary stepping stone in his train of thoughts. As mentioned earlier, it is highly likely that Baade was inspired by the works of Duyvendak, Mayall, and Oort who compiled historic data on the ancient supernova, mainly observed from Asia, in 1054, the remnant of which we today observe as the Crab Nebula. In order not to forget the astrophysical context, the general scientific aim at that time was to obtain more information about "type I" supernova events as observed in the Milky Way. These events were poorly understood in contrast to ordinary/common novae or "type II" supernovae. Since "type I" supernova events were found to be rare events, the importance of historic recordings of the sighting of "new stars" in the Milky Way was realized in retrospect and their scientific value appreciated.

W. Baade likely first started out with the acquisition of historic data of Kepler's supernova because of the accuracy of reporting and most likely because the pre-maximum brightness phase was well covered and adequately described as we shall discuss elsewhere in some more detail. However, an interesting thought is the following: the meticulous recordings of SN 1604 by Kepler and contemporary observers, just 32 years after Tycho's observations, is likely a direct consequence of Tycho Brahe's efforts (or lack of them) to take hand-written records of the steady change in brightness of the "new star" in 1572. Formulated slightly differently, the event in 1572 served as a "warning" to future observers as to do a better job (as was performed in 1604) and serves as a text-book example on how the scientific method works in practice and how it was applied early in the development of astronomy as an independent scientific research field.

The methodology developed by Baade was then applied in a similar analysis of Tycho Brahe's supernova which was lacking important data before the brightness maximum. In the following, we shall keep the chronological order and briefly report the main aspect of "type I" vs. "type II" classification.

## 5.1 Baade's 1943 paper—Kepler's nova of 1604

A few decades after the discovery and observation of SN 1572, a second nova appeared on the sky known as SN 1604 Kepler's nova. Its position, appearance, and temporal change in brightness were well documented by Kepler and contemporary astronomers. Historically, much of the experience made from observing the SN 1572 supernova directly benefited the characterization of SN 1604 (Baade, 1943).

Around the 1940s, increased attention was paid to the identification and further study of supernova remnants within the local galaxy. Baade (1943) started this endeavor by compiling historical data of SN 1604 allowing him to reconstruct the brightness variation of SN 1604.

According to the work of Baade (1943), the derived photometry was based on rough estimates. He further stated

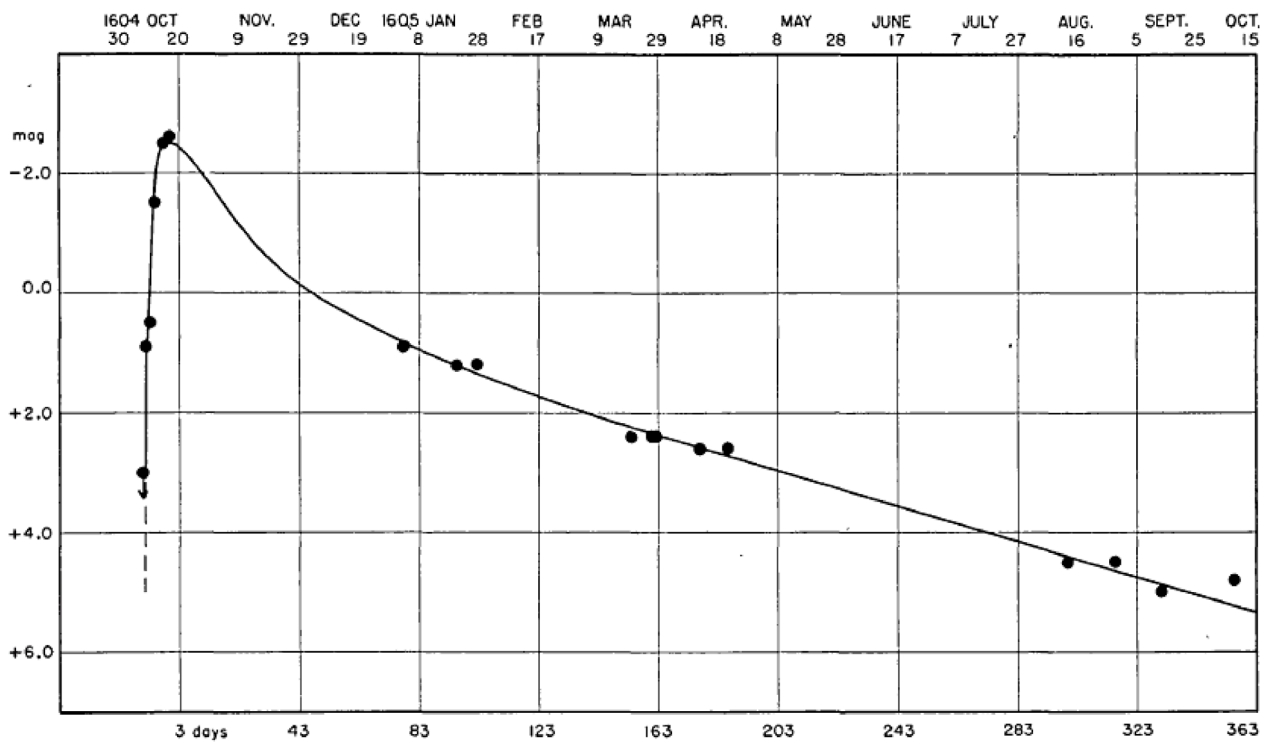

**FIGURE 12**
Light curve of SN 1604 (SN Ophiuchi) with data points derived from historical data. The solid line is the light curve of the supernova in IC 4182. © *AAS*.
The figure was reproduced with permission from the work of Baade (1943).

> "*It is a typical light-curve of a supernova of type I. If any proof is needed, it is provided by the curve in* Figure 1 [Figure 12] *representing the decline of the nova from maximum to the end of the observations. This curve is actually the visual light-curve of the recent supernova in IC 4182, properly adjusted. The remarkable agreement in the light-changes of the two stars is characteristic of supernovae of this type, which all follow closely the same pattern. Minor variations in the widths and heights of the maxima occur; but, when supernovae of this type have reached the final decline, which sets in 80–100 days after the maximum, the further decrease in brightness is the same for all, with a linear gradient of* $+0.0137 \pm 0.0012$ *mag. per day. Since the nova of 1604 conforms to this pattern, we conclude that it was a supernova of type I.*"

This is the first time where Baade (1943) classified a historic supernova as of type I. We highlight the point that no spectra were presented before classification of either type I or II based on spectral observations as proposed by Minkowski (1941). No explicit reference to the work of Minkowski (1941) was given which seems odd given that Minkowski (1941) introduced the two classes type I and type II. Rather the statement of type I classification is made based on the "Rosetta Stone" supernova IC 4182 as mentioned earlier. In the next section, we will delve into this in some more detail. However, Baade (1943) provided two references to works published by Minkowski: the first instance of citation is *i)* concerned with the faintness and interstellar absorption of the Nova Ophiuchi nebulosity (the remnant of SN 1604) where Baade wrote "*This faintness of the nebulosity in the photographic region is without doubt due to selective absorption*" and provided a reference to the work of Minkowski (1943). At the end of this paper, Minkowski wrote "*The similarity of the spectrum to that of the Crab nebula suggests that the nebulosity is the remnant of a supernova rather than that of an ordinary nova. The results thus give supplementary evidence that the nebulosity is really a remnant of Kepler's nova of 1604 and that this star was a supernova of type I.*" As a result, it is important to note that Minkowski offers a classification based on spectroscopic data of Nova Ophiuchi) and the second instance is *ii)* concerned with a comparison of spectrophotometric data of Nova Ophiuchi with the Crab Nebula where he notes similarities between the objects in their spectral characteristics where the main part of the emission between 3500 Å (blue) and 5000 Å (red) of Nova Ophiuchi is due to a continuous spectrum.

To complete this discussion of the work presented by Baade (1943), the author managed to identify the remnant of SN 1604 from a deep long-term exposure. To mitigate the effect of heavy galactic absorption, his observation made use of red filter and indeed successfully found a small diffuse patch nearby the location measured by Kepler and contemporary astronomers in 1604. Modern x-ray observations confirm this result.

## 5.2 Baade's 1945 paper—Tycho's nova of 1572

After satisfactorily classifying SN 1604, Baade embarked on compiling historical data for Tycho's supernova SN 1572. The main objective was to investigate and determine whether SN 1572 was of type I or type II. The methodology used is similar to the SN 1604 classification and is mainly based on the Rosetta Stone light curve as observed for IC 4182. Following the work of Baade (1945), he wrote

> "It has been pointed out in a previous paper1 that B Cassiopeiae [SN 1572], the bright nova of 1572, was undoubtedly a supernova because of its amplitude, which exceeded 22 mag. The recent recognition of two types of Supernovae makes it desirable to decide whether the star was a supernova of type I or type II. The light-curve of the nova, derived in the present paper, clearly indicates a supernova of type I. Because it throws new light on the final state of a supernova, B Cas is of particular interest."

Baade made some interesting remarks on the precision of the magnitude estimates for each of Tycho's observation. To achieve a mean error of 0.25 mag, Tycho must have—"*consciously or unconsciously*" (Baade, 1945)—made use of the knowledge of the observed brightness (using Ptolemy's classes or magnitude system) of the nova during a certain time period. This only makes sense given that each class in Ptolemy's system spans a significant range in magnitudes, and hence, it must have been difficult to judge the correct brightness at a given time. Making use of a time span enables the inference of a mean estimate which explains the low uncertainty of 0.25 mag. Whether or not this choice was made conscious or unconscious, it is likely one of the first applications of minimizing uncertainties for random measurements.

In his discussion on the derivation of the type of supernova, Baade (1945) provided the following peculiar statement:

> The light-curve itself is typical of a supernova of type I as shown by comparison with the visual light-curves of two other supernovae of this class, SN Oph (1604) [SN 1604] and SN in I.C. 4182 (see Figure 1) [Figure 13 in the present work]. As will be pointed out in a later paper, a very characteristic feature of supernovae of type I is the linear decrease in brightness which sets in at about 120 day after maximum and is characterized by a gradient of $+0.0137 \pm 0.0012$ mag. per day. With a gradient of $+0.014$ mag. per day for the phase interval 120 d—460 d B Cas conforms to this pattern perfectly. The descent from the maximum is less steep in B Ca than in the recent SN in I.C. 4182, but it is quite evident from the data now available that there is some variation in the form of maxima of supernovae of type I, especially in their heights and widths.

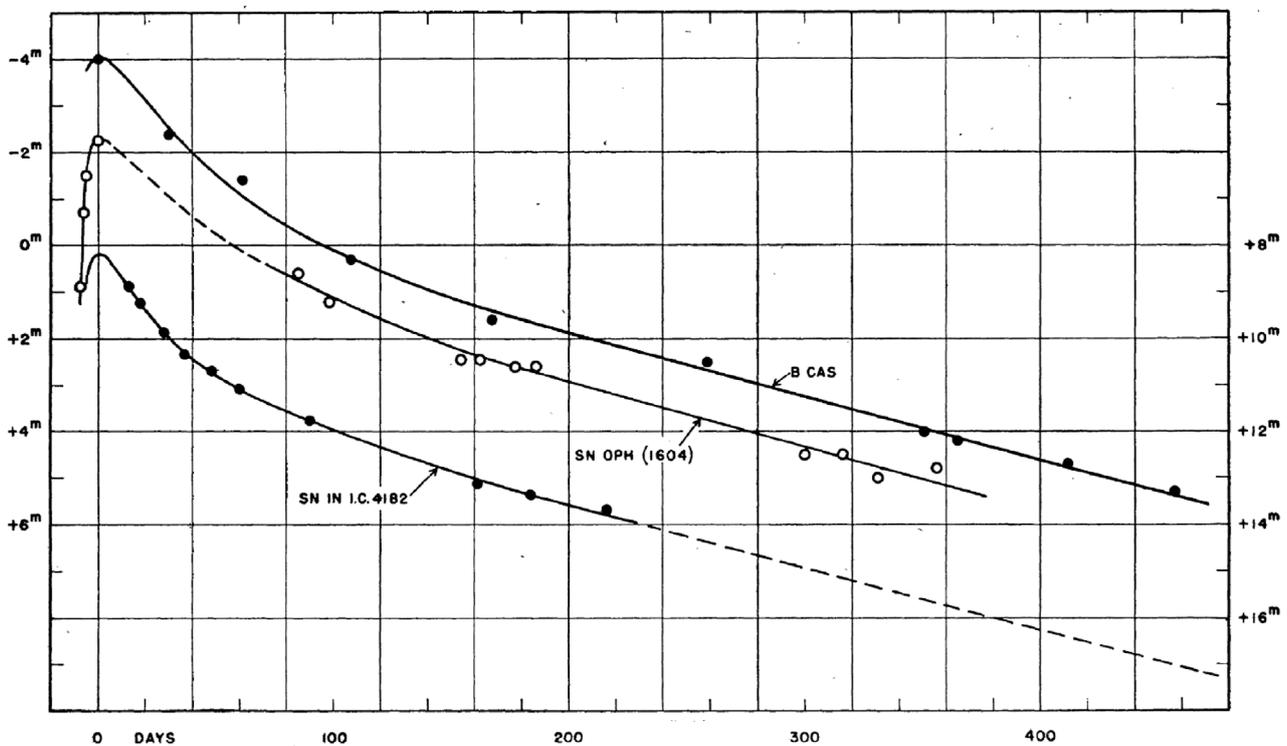

**FIGURE 13**
Light curve (top) of SN 1572 (B Cassiopeia) with data points derived from historical data as compiled by Tycho Brahe and contemporary astronomers (Baade, 1945). The bottom curve is the light curve of the Rosetta Stone supernova in IC 4182. The middle curve shows the light curve of SN 1604, another type Ia supernova, discovered by Johannes Kepler. © *AAS*. The figure was reproduced with permission from the work of Baade (1945).

Again, and as was the case for SN 1604, Baade (1945) compared Tycho's observations with the "Rosetta Stone" light curve of IC 4182 for which spectra were obtained and classified as a type I supernova. Parallel shifting of a best-fit curve to IC 4182 reveals similarities in the shape of the light curve, especially the later part of the nova's brightness decline.

Here, we have found the Holy Grail that allowed Baade to conclude that SN 1572 was a supernova of type I. The methodology applied is the same as applied for SN 1604. Baade wrote

> "The light-curve itself is typical of a supernova of type I as shown by comparison with the visual light-curves of two other supernovae of this class, SN Oph (1604) [SN 1604] and SN I.C. 4182 [IC 4182]."

The peculiar wording by Baade might suggest that the data derived from Tycho's observations were compared with two *distinctly different* supernovae. However, this is not the case. Practically, Baade compared his SN 1572 data with the light curve found for IC 4182.

To follow the logic of reasoning and as discussed earlier, we retrace the findings obtained by Baade (1943). In this paper, Baade published historic data of SN 1604. He shifted the light curve of IC 4182 vertically (in brightness) and was able to match the historical data of Kepler's (SN 1604) supernova. In particular, Baade paid special attention to the linear trend approximately 100–500 days after the maximum which is similar for the three events. From this insight, and since IC 4182 was classified as a type I supernova event (Minkowski, 1939; 1941) Baade (1943) drew the conclusion that SN 1604 must have been a type I supernova. Here, Baade applied the logic of *reasoning by analogy* to be discussed in some more detail shortly.

Finally, the same type of reasoning was also applied to SN 1572 in the work of Baade (1945) where he compares the reconstructed historic light curve of SN 1572 with SN 1604 and IC 4182. They are all similar to each other, and since IC 4182 has a type I spectrum, therefore, following the logic of reasoning by analogy, we must conclude that SN 1572 was also a type I supernova. More modern and recent studies of SN 1572 identified this historic event as a type Ia supernova (the Minkowski type I class was later renamed to type Ia) based on the observation of light echoes.

The "reasoning-by-analogy-trick" that Baade (1943) applied to SN 1604 by shifting the light curve for IC 4182 in the vertical direction might at first give rise to point at a possible source of a "flaw in logic" or the application of a false analogy. However, in this particular case, a shift in brightness (vertical shift) would physically mean a shift in distance. This means one object, in this case IC 4182, would be imaginatively moved to the distance of the other object such as SN 1604 or SN 1572. Therefore, we can now compare the two objects fairly allowing the application of the logic of reasoning by analogy. However, for this logic to work, one would have to assume that the two events IC 4182 and SN 1572 (SN 1604) must be identical prior to the supernova event in order to trace out similar evolutionary characteristics.

The aforementioned discussion was more or less pointed out by Anne Decourchelle (Decourchelle, 2017). She wrote "*The similar light curve profile of B Cas, compared to those of two type Ia SNe* (*the galactic SN Oph* [Kepler's Supernova of 1604] *and the extragalactic SN in IC 4182*), *argues for it being for a type I supernova, a characteristic feature being the linear decrease in brightness after about 120 days after maximum.*" The light curve itself is typical of a supernova of type I as shown by comparison with the visual light curves of two other supernovae of this class, SN Oph (1604) and SN in I.C. 4182.

# 6 Conclusion

In the present paper, we reviewed the early chronological developments of the field of supernova science with a focus on the historical identification of Tycho Brahe's supernova SN 1572 as a supernova of type I.

We have described how Tycho Brahe's own historical recordings from 1572 and onwards were essential in the identification and how the process coincided with the very early development of the field just prior to World War II by a handful of early supernova researchers.

However, we argue that the instrument often ascribed as the one used by Tycho Brahe in relation to his work on the supernova was in fact not used for his observation of SN 1572.

While we conclude here that the scientific importance of Tycho Brahe's recordings probably cannot be underestimated, we also note that other historical supernovae have played an important role for the initial development of supernova science, such as Kepler's supernova SN 1604 and the Chinese supernova of 1054.

Additionally, we argue that the experience and knowledge following Tycho Brahe's 1572 and subsequent observations probably led to the detailed recording of the following supernova, SN 1604 attributed to Johannes Kepler.

Furthermore, as it turns out, the supernova associated with IC 4182 plays an absolutely fundamental role in identifying the type of SN 1572 as a kind of Rosetta Stone through reasoning by analogy: as we have shown here, the early supernova researchers equated supernova peak brightness with supernova type, which allowed the identification of SN 1572 on the basis of its light curve as derived from Tycho Brahe's observations, rather than based on its unobtainable spectrum. In our review, we have demonstrated how this reasoning was part of early supernova science in the decade that followed the spectral identification of the two original types of supernovae.

## Author contributions

TH: writing-original draft. BD: conceptualization, investigation, methodology, project administration, resources, supervision, and writing-review and editing. LO: conceptualization, investigation, methodology, project administration, supervision, validation, and writing-review and editing. JH: conceptualization, investigation, methodology, supervision, validation, and writing-review and editing.

## Acknowledgments

This research has made use of NASA's Astrophysics Data System Bibliographic Services https://ui.adsabs.harvard.edu/. The idea behind this work was originally raised and proposed by Bertil Dorch. TCH acknowledges good leadership support by, and encouraging discussions with, Jens Dam during the (at times) challenging period of writing the manuscript. Also the authors would like to acknowledge Prof. M. Turatto for permission of reproducing Figure 11. The authors would like to acknowledge Prof. M. Pilar Ruiz Lapuente for fruitful discussion within the field of the early developments of research within novae and supernovae in the 1930s. Also the authors would like to express special thanks to Prof. Virginia Trimble (University of California, Irvine) for in-depth clarification on the historical development of supernova science. The first author dedicates this paper to his mom, Meike Sasse, who passed away in December 2022 - forever missed, never forgotten.

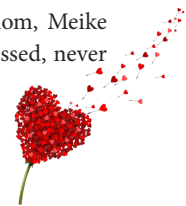

## Conflict of interest

The authors declare that the research was conducted in the absence of any commercial or financial relationships that could be construed as a potential conflict of interest.

## Publisher's note

All claims expressed in this article are solely those of the authors and do not necessarily represent those of their affiliated organizations, or those of the publisher, the editors, and the reviewers. Any product that may be evaluated in this article, or claim that may be made by its manufacturer, is not guaranteed or endorsed by the publisher.

# Appendix: Method of literature search

This review is based on a literature search using NASA's Astrophysics Data System (NASA ADS) bibliographic services (references here). The relevant bibliography was compiled searching refereed articles within the "astronomy" bibliographic collection[6]. Some initial experimentation resulted in the largest number of search results when a Boolean AND search was applied on full-text search. Therefore, this review is based on the following search command:full: "Tycho" AND full: "Brahe" AND full: "1572" AND year: 1572–2022.

This resulted in a total of 462 peer-reviewed publications in the time period from 1572 to 2022. Additional keywords such as "B Cassiopeiae" or "B Cas" or "Tycho's Star" often appeared in conjunction with "Brahe" or "1572". It seems plausible that any publication concerning the supernova of 1572 has at least the strings "Tycho," "Brahe," and "1572" in the main body of the manuscript. The experimentation yielded the insight…

Care has to be exercised when attempts are made to further reduce the number of search results. For example, the aforementioned search command could be augmented with a "AND title: '1572'" based on the assumption that all relevant publications should also contain the string "1572" in the title header. The result of adding the additional constraint would result in 23 relevant publications. This would inevitably miss the canonical paper by Baade (1945) (B Cassiopeiae as a Supernova of Type I) who appears to be the first to quantitatively translate Tycho Brahe's own written historical records and those of contemporary observers to a modern magnitude scale.

---

6  ADS maintains three bibliographic collections (totaling 15 million records) covering publications in astronomy and astrophysics, physics, and general science, including all arXiv e-prints.

Another difficulty in narrowing down the literature is the use of "type I" and "type II" in different context. Baade (1944) published a paper completely unrelated to common nova or supernovae. In this paper, Baade described two types of stellar classes (type I and type II) that were first proposed by Oort in 1936.

Searching for strings like "full: 'Tycho' AND year: 1572–2022," "abstract: 'Tycho' AND year: 1572–2022," or "title: 'Tycho' AND year: 1572–2022" or would result in 13,919, 1250, and 548 search results, respectively. Obviously, the first search method would also retain all publications related to the Tycho catalogs compiled from the Hipparcos satellite mission.
Including the criterion "AND full: '1572'" reduced the number of returned results from 3,979 to 462.

Narrowing step 2: This step excluded additional papers from judging their relevancy by the information content in their titles. For example, the title "No Surviving Companion in Kepler's Supernova" (Ruiz-Lapuenta et al., 2018; ApJ, 862, 124) or "Comets" (Barbieri and Bertini, 2017; NCimR, 40, 335) do contain the strings "Tycho", "Brahe," and "1572," but those papers are highly likely concerned with other matters not directly related to SN 1572 (Brahe, 1573).

Narrowing step 3: We only consider publications in the English language. We counted 54 search results to be written in the French language lowering the number of potential interesting papers to 408. Statistics: 231 search results have no citations to them (including two publications in 2022 and one in 2021, 2020, and 2019). Although this metric may indicate that 57% of 408 publications seem to be of less importance, we nonetheless examined all 231 abstracts for any clues of interesting results. None were found.